\documentclass{article}
\usepackage{graphicx} % Required for inserting images
\usepackage{amsmath}
\usepackage{amssymb}
\usepackage{natbib}
\usepackage{authblk}
\usepackage{xcolor}

\usepackage[margin=1in]{geometry}

\def\commenton{1}
\newcommand{\DTcomment}[1]{\if\commenton1{\color{red}(DT: #1)}\fi}
\newcommand{\KKcomment}[1]{\if\commenton1{\color{blue}(KK: #1)}\fi}
\newcommand{\HNcomment}[1]{\if\commenton1{\color{green}(HN: #1)}\fi}
\newcommand{\JKcomment}[1]{\if\commenton1{\color{teal}(JK: #1)}\fi}
\newcommand{\AVcomment}[1]{\if\commenton1{\color{purple}(AV: #1)}\fi}

\title{Evaluating Variance Estimates with Relative Efficiency}
\author[1]{Kedar Karhadkar}
\author[2]{Jack Klys}
\author[2]{Daniel Ting}
\author[2]{Artem Vorozhtsov}
\author[2]{Houssam Nassif}
\affil[1]{UCLA}
\affil[2]{Meta}
\date{}

\begin{document}
\maketitle
\section{Introduction}
Experimentation platforms in industry must often deal with customer trust issues~\cite{Radwan2024Eval}. Platforms must prove the validity of their claims as well as catch issues that arise. As a central quantity estimated by experimentation platforms, the validity of confidence intervals is of particular concern~\cite{Li2027optimalPolicies}.

To ensure confidence intervals are reliable, we must understand and diagnose when our variance estimates are biased or noisy, or when the confidence intervals may be incorrect~\cite{Weltz2023heteroskedastic}. A common method for this is \emph{A/A testing} \citep{kohavi2009controlled, kohavi2020trustworthy}, in which both the control and test arms receive the same treatment.
One can then test if the empirical false positive rate (FPR) deviates substantially from the target FPR over many tests. However, this approach turns each A/A test into a simple binary random variable. It is an inefficient estimate of the FPR as it throws away information about the magnitude of each experiment result. For instance, assuming i.i.d.\ Bernoulli draws for $n$ hypothesis tests yields a standard error of $\sqrt{(0.1)(0.9)/n}$ for the FPR, and requires around 2500 A/A tests and 5000 test groups (two to compare for each A/A test) to estimate the FPR to within 1\% with 90\% confidence.

We show how to empirically evaluate the effectiveness of statistics that monitor the variance estimates that partly dictate a platform's statistical reliability. We also show that statistics other than empirical FPR or Type I error are more effective at detecting issues. In particular, we propose a $t^2$-statistic that is more sample efficient.

\section{Measuring variance quality}
We view FPR as a \emph{variance quality metric} which can flag whether there are issues with our A/B test confidence intervals. We will also introduce two other variance quality metrics, average $t^2$ and kurtosis, commonly used for normality testing of distributions \citep{d1973tests,fisher1930moments,jarque1987test}, which we will compare against FPR.

In each A/A test, we take a collection of samples $x_{1, 1}, \cdots, x_{S, 1}$ for the control group, and $x_{1, 2}, \cdots, x_{S, 2}$
for the test group, where $S$ is the sample size for each group, and the control group and test group are both sampled from the same distribution $\mathcal{D}$. Conducting a difference of means hypothesis test, we obtain an estimate $\mu$ for difference of means, and an estimate $\sigma^2$ for the variance of the difference of means. The $t$-statistic is defined as the ratio $\frac{\mu}{\sigma \sqrt{S}}$. Across all of the hypothesis tests, we obtain a collection of $t$-statistics $t_1, \cdots, t_n$ and use these to determine the quality of our A/B tests.

For a 90 percent confidence interval, we detect a statistically significant effect for $t_j$ when
\[|t_{j}| \geq \Phi^{-1}(0.95), \]
where $\Phi$ is the inverse normal CDF. That is, we detect a statistically significant effect when the $t$-statistic lies outside of a 95 percent two-sided confidence interval.
In A/A tests, the null hypothesis is always true, so we define the \emph{false positive rate (FPR)} as the frequency at which the $t$-statistic is above this threshold:
\begin{align}\label{eq:fpr}FPR = \frac{1}{n}|\{j \in \{1, 2, \cdots, n\}: |t_j| \geq \Phi^{-1}(0.95)\}|. \end{align}
With 90 percent confidence intervals, we should expect the FPR to be close to 0.1, since it is the rate of false positives conditioned on the null hypothesis being true. To determine whether this is the case, we conduct a hypothesis test to determine whether the population FPR is equal to 0.1. Our point estimate for the population FPR is given by equation~(\ref{eq:fpr}). Observe that FPR is an average of Bernoulli random variables $Z_j$, where $Z_j = 1$ when $|t_j| \geq \Phi^{-1}(0.95)$ and $Z_j = 0$ otherwise. Under the assumption that the $Z_r$ are i.i.d., we estimate the standard error of our FPR estimate as
\[\text{StandardError}(FPR) \approx \sqrt{\frac{(FPR)(1 - FPR) }{n}}. \]
With this point estimate and standard error, we can conduct a standard $z$-test to accept or reject whether the FPR is greater than 0.1, and use the result of this hypothesis test to determine if there are mistakes in our A/B testing setup.

The false positive rate is not the only metric one can track from A/A testing. By the central limit theorem, in the limit of a large number of samples, we should expect the $t$-statistics $t_1, \cdots, t_n$ to be approximately distributed according to a standard Gaussian under the null hypothesis, so the average value of $t_j^2$ should be close to $1$. Therefore, we define the sample average $t^2$ as
\[\text{Average $t^2$} = \frac{1}{n}\sum_{j = 1}^n t_j^2. \]
Since this is an average of a collection of random variables, we again estimate its standard error as
\[\text{StandardError}(\text{Average $t^2$}) \approx \frac{\sigma_{t^2} }{\sqrt{n}},  \]
where $\sigma_{t^2}$ is the sample standard deviation. As with FPR, we conduct a hypothesis test to determine whether the average $t^2$ is equal to $1$.

The third metric we consider is the \emph{kurtosis} of the $t$-statistics. For a random variable $X$, its kurtosis is defined as its fourth standardized moment $\mathbb{E}[(X - \mu)^4/\sigma^4]$, where $\mu$ and $\sigma$ are the mean and standard deviation of $X$. A standard Gaussian random variable has zero kurtosis. So if the $t$-statistics $t_1, \cdots, t_n$ are distributed according to a standard Gaussian, we should expect the kurtosis of their distribution to be close to zero. The standard unbiased estimator of the kurtosis of the $t$-statistics is
\[g_2 = \frac{(n - 1)}{(n - 2)(n - 3)}\left[(n + 1)\frac{M_4}{M_2^2} -3 (n - 1) \right], \]
where $M_2$ and $M_4$ are the second and fourth sample moments of the $t$-statistics about the mean respectively. With a normal approximation, the standard deviation of this estimator is approximately
\[\sqrt{\frac{24n(n - 1)^2}{(n - 3)(n - 2)(n + 3)(n + 5)}}  \]
(see~\citep{fisher1930moments} for details). As with FPR and average $t^2$, we can use this standard deviation estimate to conduct a hypothesis test for kurtosis.

\section{Comparing methods}
Given these different variance quality metrics, we will consider their efficiency in the context of a particular procedure. Suppose that the control and test samples are drawn from a probability distribution $\mathcal{D}$. We conduct A/A tests as mentioned above, producing a set of means and variances of the lifts $\{(\mu_j, \sigma^2_j)\}_j$. Suppose that we instead observe noisy versions of the variances $\hat{\sigma}^2_j = \sigma^2_j \xi_j$, where the $\xi_j$ are sampled i.i.d.\ from a noise distribution. We would like the variance quality metrics to detect that the variance estimates are noisy. Hence, we define the \emph{power} $1 - \beta$ of a variance quality metric as the probability that it rejects its null hypothesis in the presence of this noise. We define the \emph{sample complexity} $N(\alpha, \beta)$ for a given variance quality metric as the number of permutations necessary to attain a given power $1 - \beta$ while using significance level $\alpha$ for its hypothesis tests. We define the \emph{relative efficiency} of metric $1$ with respect to metric $2$ as the ratio of their sample complexities:
\[e_{12} = \frac{N_2(\alpha, \beta)}{N_1(\alpha, \beta)}. \]

This is a finite-sample version of Pitman efficiency \cite{pitman1949notes}. We will determine the relative efficiency of the three variance quality metrics in this noisy A/A test setting.

\section{Experiments}
\begin{figure}[h]
    \centering
    \includegraphics[width=0.8\linewidth]{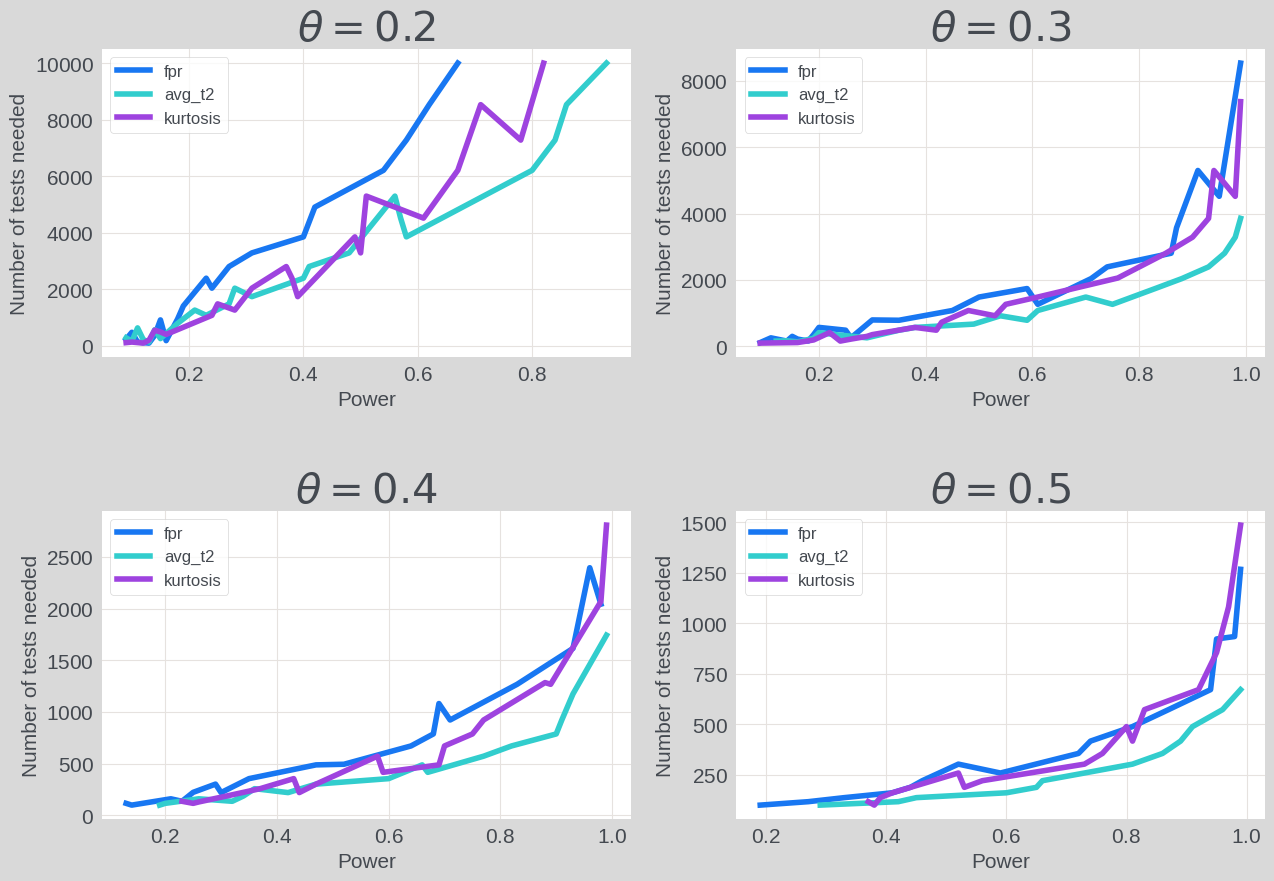}        \caption{Sample complexity of attaining a given level of power for different tests of variance quality.}
    \label{fig:sample-complexity}
\end{figure}

We conduct experiments with the following setup. We take $S = 1000$ samples for each control and test group drawn from a uniform distribution on the range $[5, 6]$. We take our noise distribution to be parameterized by a value $\theta$, and take the family of noise distributions to be lognormal distributions with parameters $\mu = -\frac{\theta^2}{2}$ and $\sigma = \theta$. This noise is multiplied to the true variance and is unbiased (has expectation 1). All of our hypothesis tests use significance level $\alpha = 0.1$. To calculate the power $1 - \beta$, we compute the frequency that each hypothesis test detected a statistically significant result over $500$ trials. We run trials over a sweep of values for the number of hypothesis tests $n$ and the noise level $\theta$. For $n$, we take 30 values between $10^2$ and $10^{4}$ uniformly spaced in log-space. For $\theta$, we take the values $\{0.1, 0.2, 0.3, 0.4\}$. Thus, the result of each experiment can be expressed as a triple $(\theta, n,1 -  \beta)$, where $\theta$ is the noise level, $n$ is the number of hypothesis tests, and $1 - \beta$ is the resulting power.
To judge the sample complexity, we plot the pairs $(1 - \beta, n)$ for each value of $\theta$ and record the results in Figure~\ref{fig:sample-complexity}.

\begin{figure}[h]
    \centering
    \includegraphics[width=0.95\linewidth]{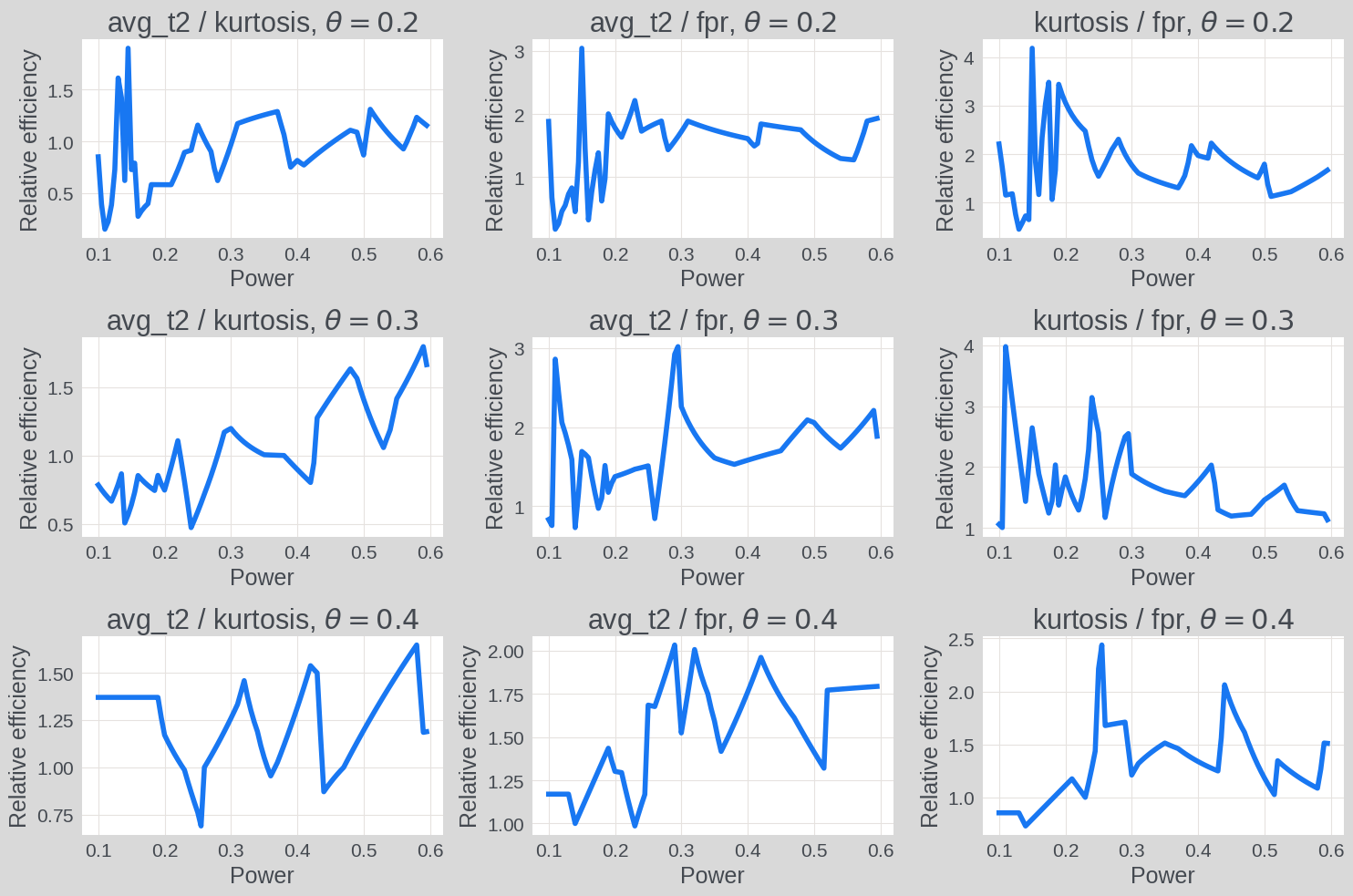}
    \caption{Pairwise evaluations of relative efficiency over different parameter values of the noise distribution. }
    \label{fig:relative-efficiency}
\end{figure}

In the first case of $\theta = 0.2$, each of the variance quality metrics requires several thousand hypothesis tests to attain sufficient power.  In the other three cases, each of the metrics is able to detect the noise within a few thousand hypothesis tests. We see that average $t^2$ is consistently more sample-efficient than FPR, and the same is generally true for kurtosis to a lesser extent (see Figure~\ref{fig:sample-complexity}).

We further track this by linearly interpolating the sample complexity curves above, and then computing the relative efficiency for each pair of metrics. We record the results for $\theta \in \{0.2, 0.3, 0.4\}$. The sample complexity plot reveals that average $t^2$ has a relative efficiency around $1.5$ to $2$ compared against FPR (Figure~\ref{fig:relative-efficiency}). The improvement is greater and more consistent for higher values of power.
These two experiments show that both average $t^2$ and kurtosis outperform FPR in the presence of ``unbiased" noise in the variance estimate.

\section{Conclusion}
In this work, we proposed a framework for evaluating variance estimates for A/B test confidence intervals using variance quality metrics which are functions of $t$-statistics from A/A tests. We compare the variance quality metrics using a finite-sample notion of relative efficiency. Comparing average $t^2$ and kurtosis, we find that both are 1.5 to 2 times more sample-efficient than FPR, with average $t^2$ achieving this improvement more consistently.

\bibliography{refs}
\bibliographystyle{plain}
\end{document}